\documentclass[12pt]{article}

\usepackage{a4wide}
\usepackage[pdftex,usenames,dvipsnames]{color}
\usepackage{graphics}
\usepackage{amsfonts}
\usepackage{amssymb}
\usepackage{accents}

\newcommand{\nit}{\noindent}

\newcommand{\vs}[1]{\vspace{#1 ex}}
\newcommand{\hs}[1]{\hspace{#1 em}}
\newcommand{\bfr}{\begin{flushright}}
\newcommand{\efr}{\end{flushright}}
\newcommand{\bc}{\begin{center}}
\newcommand{\ec}{\end{center}}
\newcommand{\ben}{\begin{enumerate}}
\newcommand{\een}{\end{enumerate}}

\newcommand{\be}{\begin{equation}}
\newcommand{\ee}{\end{equation}}
\newcommand{\ba}{\begin{array}}
\newcommand{\ea}{\end{array}}

\newcommand{\Del}{\Delta}

\newcommand{\brv}{\bar{v}}

\newcommand{\lh}{\left(}
\newcommand{\rh}{\right)}

\begin{document}

\pagestyle{plain}
\pagenumbering{arabic}

\nit
\bc
{\bf On Huygens' derivation of the laws of elastic collisions}
\vs{2}

Jan-Willem van Holten 
\vs{1}

Lorentz Institute, Leiden University, \\
and \\
Nikhef, Amsterdam, NL 
\vs{2}

November 2025
\ec
\vs{3}

\nit
{\small
{\bf Abstract }\\
In this note I sketch the work of Christiaan Huygens to develop a theory of motion and its 
application to elastic collisions. In this theory he uses the relativity of uniform linear motion 
to  derive the conservation of momentum and kinetic energy (at the time referred to as 
{\em living force} or {\em vis viva}). The conservation of living force was used subsequently 
by Leibniz as a basic general principle of dynamics, an alternative to that of Newton set forth 
in the {\em Principia Mathematica}.
}
\vs{2}

\nit
{\bf 1.\ Background}

\nit 
Christiaan Huygens (1629-1695) was the second son of Constantijn Huygens sr.\ and Suzanna van Baerle. His 
elder brother was Constantijn jr., who became secretary of William III, Prince of Orange, Stadholder of Holland 
and as of 1689  King of the United Kingdom of England, Scotland and Ireland. His father, a high-ranking diplomat, 
was acquainted with Ren\'{e} Descartes, who lived in Holland and at times visited the Huygens' home Hofwijck in 
Voorburg near The Hague.\footnote{ For a biography of Chr.\ Huygens see: H.\ Aldersey-Williams, 
{\em Dutch Light} (Picador, 2020); \\  C.D.\ Andriessen, {\em Titan}  (Utrecht University, 2003).}

Christiaan studied law at the University of Leiden, but spend more time on mathematics with his teacher Frans 
van Schooten, who knew and admired Descartes. Christaan also closely studied Descartes' work on 
mathematics and physics and was influenced by him, especially in his early years. His numerous discoveries 
in mathematics, physics and  astronomy made Huygens the most prominent scientist of his generation 
in Europe. As such he was appointed by the French minister of state Colbert to become the first scientific 
director of the Royal Academy of Sciences in Paris. 

In view of his extensive use of new mathematical techniques in algebra and geometry, such as introduced by 
Descartes and Vi\`{e}te, Christiaan Huygens is often considered as founder of theoretical physics. One of 
the first problems Huygens as a young man set out to investigate was the theory of motion, taking Descartes' 
work as his starting point. Descartes had developed the idea that the world consisted of matter in motion and 
was governed by fixed laws of nature. He thereby rejected the existence of supernatural forces and the 
necessity of divine interventions in the workings of the universe. Based on these convictions he developed 
a number of ideas on the motion of material bodies which could be formulated mathematically.\footnote{
R.\ Descartes. {\em Principia Philosophiae} (Elzevier, Amsterdam; 1644).}

First he introduced  a law of conservation of motion, necessary to prevent the world from running down or flying 
apart, without which divine intervention would indeed become necessary to preserve its ordered existence. Next
he stated a number of laws of the collisions of moving bodies he thought would allow a theory of space filled
with matter of different degrees of fineness, the constant collisions of which were the cause of the observed 
motions of bodies and the changes in their configurations. For example, vortices in this fine matter might explain 
the motion of planets around the sun and of moons around their planets. 

It is not necessary here to present a detailed list of Descartes' laws of motion. Suffice to mention that his law 
of conservation of motion was stated as the total amount of speed weighted by matter being fixed:
\[
\sum_i m_i v_i = \mbox{constant},
\]
$m_i$ denoting the amount of matter in body $i$, and $v_i$ its speed being taking as a magnitude, not as a 
vector quantity specified in addition by its direction; as an example of his rules of colliding bodies, his fourth rule 
states that if a smaller body collided with a larger one at rest, it would bounce back without transferring any of its 
motion to the larger body, which would therefore stay at rest. 

Huygens soon discovered that these rules were inconsistent and did not describe the observed effects in
elastic collisions of hard spherical bodies (his terminology). He replaced Descartes' laws with his own 
theory of motion, mathematically consistent and in better agreement with observations. On the basis of a 
few simple rules he was able to derive the conservation of momentum and energy, which served as starting 
point for Newton's laws of motion\footnote{I.\ Newton, {\em Principia Mathematica} (1687)} and Leibniz's theory 
of living force.\footnote{G.W.\ Leibniz, {\em Brevis demonstratio}, Acta Eruditorum, 5 (1686), 161-163.} These results were already largely 
included in his notes from 1652 and in an unpublished manuscript of 1656. He eventually made it into a 
contribution to the proceedings of the Royal Society in London and published them in the French Journal 
des S\c{c}avans in 1669.\footnote{Chr.\ Huygens, {\em Lettre}, Journal des S\c{c}avans (March 18, 1669), 
22; \\ {\em De motu corporum ex mutuo impulsu}, Philosophical Transactions Vol.\ IV (April 12, 1669), 925.} 
The theory was published in extended form posthumously in the latin version {\em De Motu Corporum ex 
Percussione} in 1703.\footnote{Chr.\ Huygens, {\em Oeuvres compl\`{e}tes}, Tome XVI, ed.\ J.A.\ Volgraff; 
(Nijhoff, Den Haag, 1929). }

Being critical of Descartes' theory of motion, Huygens nevertheless did adhere to his idea of space filled
with an ether of fine matter through which motion could be transferred by contact interactions, to provide a 
mechanical explation of phenomena like gravity and light. He considered his own work as laying an improved
foundation for such a mechanical world view. 
\vs{1}

\nit
{\bf 2.\ Huygens's theory of motion}

\nit
Huygens' theory of motion started from a few simple principles. The first one is the law of inertia: that 
{\em all bodies once in motion in the absence of obstacles continue to move with constant speed along a 
straight line}. Huygens was indeed the first to state this law in its correct form.\footnote{For more 
on the history and background of this principle see:  E.J.\ Dijksterhuis, {\em The mechanization of the world 
picture} (Oxford Clarendon Press, 1961).} 

Next he stated his basic law of elastic collisions of spherical bodies: that {\em two equal bodies, moving straight 
towards each other with equal speed in opposite directions, after colliding bounce back with the same equal 
speed they had before}. This law combines the symmetry of the event with the conservation of motion.

Finally he introduced the principle of relativity: that {\em the motion of bodies is to be understood with respect 
to others considered at rest, irrespective of whether the first and the latter bodies are subject to a common
motion}. In the next sentence he specified that such common motion was to be taken as a {\em uniform} 
motion. Clearly from a modern perspective these three laws apply only to motion in inertial frames, and the 
relative motion of inertial frames is necessarily constrained to be uniform. 

A first consequence of these principles Huygens deduced was, that if two equal bodies move straight 
towards each other with different velocities, after the collision they would have their velocities interchanged. 
For equal velocities this result reduces to Huygens' second rule; that it holds also for unequal velocities
follows directly from the principle of relativity: one can consider the collision in a different frame in which the 
velocities of the two bodies are equal and opposite. Indeed, if the two different velocities are $v_1$ and 
$v_2$, this is achieved by observing the event from a frame moving with velocity $(v_1 + v_2)/2$ with
respect to the original frame of reference. In this frame the velocities are $\brv_1 = - \brv_2 = (v_1 - v_2)/2$.
As the result holds in this frame, and it must also hold in the original one.

Next Huygens generalizes this result to arbitrary bodies by positing that, when they approach each other 
along a straight line with relative velocity $u = v_2 - v_1$, after a collision they continue with the opposite 
relative velocity $v'_2 - v'_1 = -u$. For equal bodies this is obvious from the previous result. For unequal 
bodies he could have introduced it as a new fundamental principle, but instead he choses to derive it from 
another one: that if one of the two masses is observed after a collision to reverse its velocity (without change 
of magnitude) the other mass also reverses its velocity. Starting from the situation in which the body $m_2$ 
is initially at rest and moves away with velocity $v_2' = u$ after the collision, this same body is seen to 
reverse its motion in a frame with velocity $u/2$ with respect to the original one. In this frame the velocities 
of $m_2$ before and after the collision are $\brv_2 = - u/2$ and $\brv'_2 = u/2$. By hypothesis in this frame 
then also $\brv'_1 = - \brv_1$, and therefore $\brv'_2 - \brv'_1 = - (\brv_2 - \brv_1)$. In any other frame 
moving with constant velocity to the original rest frame of $m_2$ all velocities are changed by a fixed 
amount; as a result their differences remain the same and the forgoing conclusion holds in all frames.  

Having established this result Huygens now makes a remarkable move; he imagines the colliding bodies 
gaining their speed $v$ from being deflected in the horizontal direction after a free fall from a height $h$, 
using Galileo's law $v^2 = 2 g h$. After the collision they can then rise up again to the new height 
$h' = v^{\prime\,2}/2g$. Now noting the reversibility of the motion of the bodies and his own generalization 
of Toricelli's axiom, implying that the center of gravity cannot without external intervention rise to a greater 
height than where it started from rest, he argues that the center of gravity must return to the same height 
after the collision. This then implies the equality 
\be
m_1 v_1^2 + m_2 v_2^2 = m_1 v_1^{\prime\,2} + m_2 v_2^{\prime\,2},
\label{1}
\ee
for motion before and after the collision. It expresses the conservation of what Leibniz would later call 
{\em living force}, equivalent to kinetic energy in present terminology.  

Combining this result with the previous one, the conservation of momentum can be derived 
straightforwardly. First 
\[
m_1 \lh v_1^{\prime\,2} - v_1^2 \rh = - m_2 \lh v_2^{\prime\,2} - v_2^2 \rh \hs{1} \Leftrightarrow \hs{1}
m_1 (v'_1 - v_1) (v'_1 + v_1) = - m_2 (v'_2 - v_2)(v'_2 + v_2)
\]
It is easily seen that the above rule of reversing the relative velocity of the bodies in the collision is 
equivalent to
\[
v'_2 + v_2 = v'_1 + v_1.
\]
Dividing by this common factor and reshuffling the terms of the equation one finds
\be
m_1 v'_1 + m_2 v'_2 = m_1 v_1 + m_2 v_2.
\label{2}
\ee
Although Huygens does not mention this result in the posthumously published work {\em De Motu},
it is listed explicitly, together with the conservation of living force, as a characteristic of the collision 
of hard bodies in his paper of 1669 in the Journal des S\c{c}avans. 

By turning around the above argument it is easily seen that Huygens' rule, that in elastic collisions the 
relative velocity of the bodies is reversed, becomes equivalent to the conservation of linear momentum. 
This in turn expresses Newton's third law of the equality of action and reaction as applied to elastic 
collisions: 
\be
\Del p_1 = m_1 (v'_1 - v_1) = -m_2 (v'_2 - v_2) = - \Del p_2.
\label{3}
\ee
 \vs{1}
 
\nit
{\bf 3.\  More on conservation laws}

\nit
By definition the vertical height of the center of gravity is 
\[
h_c = \frac{m_1 h_1 + m_2 h_2}{m_1 + m_2}. 
\]
Therefore the heights of the centers of gravity before and after the collision are equal: $h'_c = h_c$, 
provided
\[
m_1 h'_1 + m_2 h'_2 = m_1 h_1 + m_2 h_2.
\]
The conservation of living force, eq.\ (\ref{1}), follows immediately by Galileo's law of free fall. 

If the conservation of living force holds in one frame, it holds in all frames in uniform relative motion, 
as shifting all velocities by a common amount $u$ such that $\brv = v - u$ results in
\[
m_1 \brv_1^{\prime\,2} + m_2 \brv_2^{\prime\,2} + 2u \left[ m_1 \brv'_1 + m_2 \brv'_2 \right] = 
 m_1 \brv_1^2 + m_2 \brv_2^2 + 2u \left[ m_1 \brv_1 + m_2 \brv_2 \right].
\]
The total linear momentum before and after collision being shifted by a constant amount 
$(m_1 + m_2)u$, its conservation is guaranteed:
\[
m_1 \brv'_1 + m_2 \brv'_2 = m_1 \brv_1 + m_2 \brv_2.
\]
This establishes the conservation of living force for arbitrary $u$.

Other results obtained by Huygens in {\em De Motu} are, that in any frame the center of mass continues 
to move with the same velocity before and after the collision. Therefore the frame in which both masses 
reverse their velocities is the frame in which the center of mass is at rest. By definition the center of
mass of the bodies $m_1$ and $m_2$ moving on a straight line with positions $x_1(t)$ and $x_2(t)$ is
the point $x_{CM}(t)$ such that
\[
m_1 \lh x_{CM} - x_1 \rh = m_2 \lh x_2 - x_{CM} \rh \hs{1} \Rightarrow \hs{1}
x_{CM} = \frac{m_1 x_1 + m_2 x_2}{m_1 + m_2}.
\]As the two bodies move with constant velocities $v_1$ and $v_2$ the center of mass also moves 
with a constant velocity, given by
\[
v_{CM} = \frac{m_1 v_1 + m_2 v_2}{m_1 + m_2}.
\]
The total momentum being conserved in the collision it follows immediately that the center of mass 
continues as before: $v'_{CM} = v_{CM}$, in any frame. Moreover, in a frame in which $v'_1 = - v_1$ 
and $v'_2 = - v_2$ it also follows that $v'_{CM} = - v_{CM}$, which can be true only if in that frame 
$v_{CM } = v'_{CM} = 0$. 
\vs{1}

\nit
{\bf 4.\  Summary and discussion}.  

\nit
The starting point of Huygens' theory of motion is his principle of inertia: 
in the absence of friction or obstacles bodies move uniformly with constant velocity along straight lines. 
In addition he holds that motion cannot be created out of nothing, neither can it disappear spontaneously, 
either process implying the other by reversal of motion. Finally Huygens firmly holds on to the principle
of relativity, the description of motion in frames moving uniformly with respect to each other being fully 
equivalent; this becomes a powerful tool in his analysis of the interactions of material bodies. Implicitly 
it denies the existence of an absolute space and time. 

For Huygens all forces between and changes in material bodies have a mechanical origin, being 
reducible to contact interactions: they result from the collisions of particles. He establishes a complete 
theory of the collisions of hard bodies, i.e.\ bodies which collide without loss of motion. The rules that 
govern such collisions imply that the total momentum (quantity of motion) and the total kinetic energy 
(Leibniz' living force) are conserved, and the center of mass of the bodies continues to move with the 
same velocity. 

These rules, including the principle of inertia, are all incorporated in Newton's laws of motion. However there 
are clear differences in their approaches. Newton's theory of motion allows for non-contact forces: forces  
at a distance like gravity, which in Huygens' view (inspired by Descartes) are impermissible. On the other 
hand, in contrast to Huygens Newton defends the necessity of absolute space and time, violating the principle 
of relativity. 

After Newton published his theory of gravity in the {\em Principia}, there were several attempts to reconcile
this theory of action at a distance with the mechanistic principles of Huygens and Descartes, continuing till 
well into the 19th century. These include for example the works of Fatio de Duillier, Le Sage and Lord Kelvin. 
All these theories have been superseded by Einstein's theory of General Relativity. Interestingly, in a public  
lecture at Leiden University,\footnote{A.\ Einstein, {\em \"{A}ther und Relativit\"{a}tstheorie}; 
inaugural address; Leiden University (May 5, 1920).} Einstein himself suggested to identify space-time itself, 
having been endowed with mechanical properties, with the ether explaining the origin of gravitational (and 
possibly other) forces.

\end{document}